\documentclass[twocolumn,prc,showpacs,showkeys,preprintnumbers,unsortedaddress,amsmath,amssymb,floatfix]{revtex4}
\input epsf.sty
\usepackage{graphicx}
\usepackage{bm}

\begin{document}
\title{\bf Screening Effects in Superfluid Nuclear and Neutron Matter within Brueckner Theory}

\author{L. G. Cao$^{1}$, U. Lombardo$^{1,2}$, P. Schuck$^{3}$}

\affiliation{$^{1}$Laboratori Nazionali del Sud, INFN, Via Santa
Sofia 62, I-95123 Catania, Italy }

\affiliation{$^{2}$Dipartimento di Fisica dell'Universit\`a, Viale
Andrea Doria 6, I-95123 Catania, Italy}

\affiliation{$^{3}$Institut de Physique Nucl\'{e}aire,
Universit\'{e} Paris-Sudd,F-91406 Orsay Cedex, France}

\date{ \today}

\begin{abstract}
Effects of medium polarization are studied for $^1S_0$ pairing in
neutron and nuclear matter. The screening potential is calculated
in the RPA limit, suitably renormalized to cure the low density
mechanical instability of nuclear matter. The selfenergy
corrections are consistently included resulting in a strong
depletion of the Fermi surface. All medium effects are calculated
based on the Brueckner theory. The  $^1S_0$ gap is determined from
the generalized gap equation. The selfenergy corrections always
lead to a quenching of the gap, which is enhanced by the screening
effect of the pairing potential in neutron matter, whereas it is
almost completely compensated by the antiscreening effect in
nuclear matter.
\end{abstract}

 \maketitle

\section{Introduction}
A satisfactory description of superfluidity in nuclear matter has
not yet been achieved despite almost fifty years of research have
elapsed since the first application of the BCS theory to nuclear
systems \cite{cms}. Somewhat at variance with the electron pairing
in superconductors the pairing in nuclear  systems results from
the interplay between the direct action of the bare nuclear force
and the action induced by the medium polarization. The attractive
components of the bare nuclear interaction have led to the
investigation of  several pairing configurations, e.g.
neutron-neutron or proton-proton pairing in the $^1S_0$ channel in
neutron stars \cite{alpar} disregarding possible repulsive effect
exerted by screening of the force via the medium. A pairing
suppression has in fact been found by most calculations of pairing
in neutron matter (see Ref.~\cite{schul} and references therein).
On the contrary, other pairing configurations have not yet been
explored since the repulsive components of the direct nuclear
interaction cannot support the formation of Cooper pairs. But
there are strong indications that, in a nuclear rather than
neutron matter environment, the medium polarization of the
interaction can favor the formation of Cooper pairs similar to the
lattice vibrations in ordinary superconductors. These indications
come both from nuclear matter calculations  and from finite
nuclei. In nuclear matter the medium enhancement of
neutron-neutron $^1S_0$ pairing is to be traced back to the proton
particle-hole excitations \cite{shen}, and in finite nuclei to the
surface vibrations \cite{milan}.

Another distinctive feature of the nuclear environment is the
presence of strong short range correlations that induce two
effects relevant for the pairing: one is the depletion of the
Fermi surface, which is experimentally supported by measurements
of electron scattering on $^{208}Pb$ \cite{nikhef}, the other one
is the strong mass renormalization caused by short-range
particle-particle correlations \cite{maha}.

The two effect conspire against the pairing formation: The
depletion of the Fermi sea reduces the phase space available for
particle-particle virtual transitions around the Fermi surface,
the mass renormalization enhances the dispersive effect of the
mean field \cite{self}.

Therefore a complete microscopic treatment of the very subtle
pairing problem requires vertex and  selfenergy corrections to be
treated and to be considered on the same footing. In a previous
paper\cite{shen} we made a study of these in-medium effects under
several simplifying assumptions. First came the approximation to
replace the Born term of the pair interaction in the S=0, T=1
channel by the Gogny force \cite{Gogny}. Though in that channel
the Gogny force is not dissimilar to the action of the bare force
( see Ref. \cite{Sed}), it shows a little too much attraction for
momenta characterizing saturation. The first improvement in the
present work is then the use of a realistic two body force (V18
\cite{v18}, see below) in the Born term. Secondly we will use as
vertices in the induced interaction a force which is based on a
more modern G-matrix calculation \cite{eos} as this was the case
for the Gogny force \cite{Gogny}. Thirdly we corrected an
unfortunate phase error which slipped into the evaluation of the
induced force at least for the symmetric nuclear matter case in
the S=0, T=1 channel what gave raise to a too strong
anti-screening effect. With these corrections and improvements we
now get reasonable renormalization effects of the pairing force
and we calculate the corresponding gaps as a function of density
in pure neutron matter as well as in symmetric nuclear matter.

In detail the paper is organized as follows. In Sec. II the
generalized gap equation is reviewed along with the approximations
on the pairing potential and selfenergy, which lead to the
determination of the energy gap. In Sec. III the screening
interaction is discussed in the RPA limit, and then the summation
of bubble diagrams and the resummation of dressed bubble diagrams
both using the Landau parameters are derived. In Sec. IV the
results are presented: first, for the separate contributions of
particle-hole (ph) scalar, vector and isovector excitations in
neutron matter and nuclear matter; second, the solution of the gap
equation for $^1S_0$ pairing with a discussion of the effects of
the selfenergy corrections and medium polarization potential.
Section V is devoted to the comparison with other calculations and
to the conclusions.

\section{Generalized gap equation}
\begin{figure}[htbp]
\vglue -1.5cm
\begin{center}
\includegraphics[width=85mm]{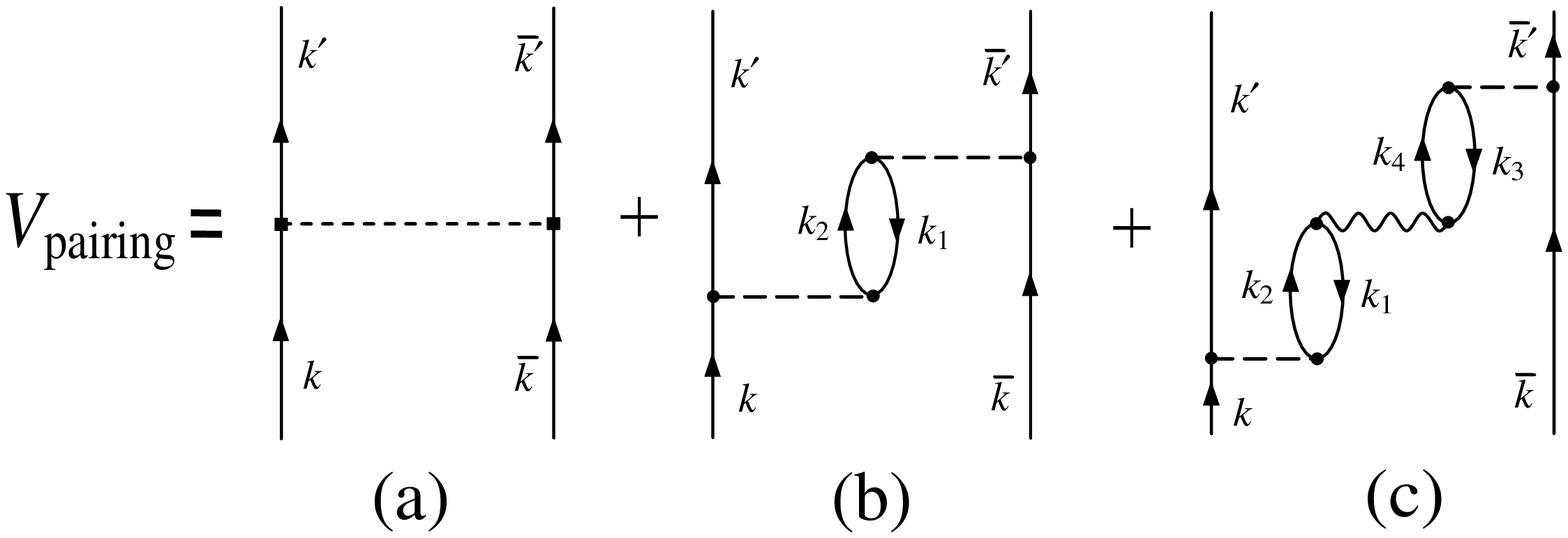}
\vglue -5.5cm  \label{f:screen}
\end{center}
\caption{Pairing interaction with screening in the RPA
approximation. The short-dashed line represents the bare
interaction, long-dashed lines the G-matrix, the wiggly line the
p-h residual interaction resummed to all orders. All vertices are
to be understood as anti-symmetrised matrix elements. The
notations are: $k = (\vec k,\sigma,\tau), \bar k =(-\vec
k,-\sigma,\tau)$.} \label{diagram}
\end{figure}

The spectrum of a superfluid homogeneous Fermi system  is derived
from the generalized gap equation\cite{nozi,mig,risch}:
\begin{equation}
\Delta_{\bm{k}}(\omega)=\sum\limits_{k^{\prime}}\int\frac{d\omega^{\prime}%
}{2\pi i}\mathcal{V}_{\bm{k},\bm{k}^{\prime}}(\omega,\omega^{\prime}%
)F_{\bm{k}^{\prime}}(\omega^{\prime}),
\end{equation}
where $\mathcal{V}$ is the sum of all irreducible NN interaction
diagrams and $F_{k}(\omega)$ is the anomalous propagator. The
class of diagrams selected for the present calculation is plotted
in Fig.~\ref{diagram}.

In nuclear matter $\mathcal{V}$ can be approximated by the bare
interaction $\mathcal{V}_0$(diagram (a)), which is responsible of
the formation of Cooper pairs, and the class of bubble insertions,
which play the role of screening. In turn, the screening
interaction $\mathcal{V}_1$ can be split into two parts: the
one-bubble term (diagram (b)) containing only the mixed
configuration with particle-particle (pp) plus ph excitations, the
multi-bubble term also containing all insertions of pure ph
interaction vertices (diagram (c)). The splitting is a convenient
way to point out that mixed vertices and pure ph vertices have to
be treated on different footing, as discussed in Ref. \cite{shen}.
The first bubble diagram can also be seen as the lowest order
correction to the Born term. As discussed afterwards, the vertex
insertions in diagrams (b) and (c) are described  by a Brueckner
G-matrix.

Within the Brueckner theory of nuclear matter, the single particle
(sp) energy spectrum of the non superfluid state is derived from
the hole-line expansion of the mass operator. The BHF
approximation, extended to include depletion of the Fermi surface
due to the strong ground state correlations, gives an important
quenching of the pairing gap and hence it can not be neglected.
The pp correlations on the pairing interaction are embodied in the
gap equation itself \cite{cms}. Consistently, the interaction
vertices in the screening term should be described in terms of the
G-matrix. In addition, the latter must be dressed according to the
Babu-Brown theory of the induced interaction \cite{babu,sjob} to
avoid the low density instability problem of nuclear matter, as
discussed later. Since the exact resummation of the bubble series
(Bethe-Salpeter equation) with G-matrix is a prohibitive task, the
ph vertex insertions can be conveniently evaluated in the Landau
limit, and eventually replaced by the Landau parameters.

 The effects of the selfenergy corrections have intensely been studied; in particular the depletion of the Fermi
surface is expected to hinder the virtual transitions around the
Fermi surface and thus its effect is to weaken the pairing
correlations. It is of particular interest to understand to what
extent the resulting quenching of pairing gap is compensated by in
medium vertex corrections that in nuclear matter are strongly
attractive. On the other hand, in the case of neutron matter
selfenergy effects enhance quenching due to screening at variance
with the predictions of recent Monte Carlo many body calculations
\cite{fantoni}.

Going beyond the pure BHF approximation, the main dispersive
corrections arise from energy dependence of the self-energy, as
shown in previous papers. The correction to the non superfluid
propagator $G_k(\omega)$ is simply a renormalization factor of its
pole part. This factor, named Z-factor, is
\begin{equation}
{\cal Z}^{-1} = 1 -
[\frac{\partial\Sigma_k(\omega)}{\partial\omega}]_{\omega=\omega_k},
\end{equation}
which measures the discontinuity of the occupation probability
around the Fermi energy. Correspondingly, the abnormal propagator
appearing in the gap equation (Eq. (1))
\begin{equation}
F_{\bm{k}}(\omega)=\frac{\Delta_k(\omega)}{G^{-1}_k(\omega)G^{-1}_k(-\omega)+\Delta^2_k(\omega)},
\end{equation}
is renormalized by a factor $Z^2$. For a static interaction the
gap function is also independent of energy and the energy
integration can be performed analytically. Since the analytical
structure of the abnormal propagator is not modified, the gap
equation takes the same form as in the pure BCS case. One easily
obtains:
\begin{equation}
\Delta _k=-\frac{1}{2}\int d^3{\vec k^{\prime}} {{\cal
V}}_{k,k^{\prime }} \frac{Z_k Z_{k^{\prime}}\Delta
_{k^{\prime}}}{\sqrt{(\varepsilon_{k'}-\varepsilon_F)^2+\Delta
_{k'}^2 }}, \label{e:zzgap}
\end{equation}
where $\varepsilon_F$ is the Fermi energy and $\varepsilon_k$ is
the on-shell self-energy. The preceding gap equation is equivalent
to the BCS version except for the $Z^2$ factor which is modeling
the effect of the interaction around the Fermi surface. Since the
value of the Z-factor, though depending on the ground state
correlations, is always less than unity in the vicinity of the
Fermi surface, inevitably the energy gap will turn out quenched in
this respect. Our predictions of the gap in nuclear and neutron
matter presented in this paper rely on the solution of the latter
equation.

\section{Screening interaction}
\subsection{One-bubble screening interaction}

\begin{figure}[hbtp]
\includegraphics[scale=0.40]{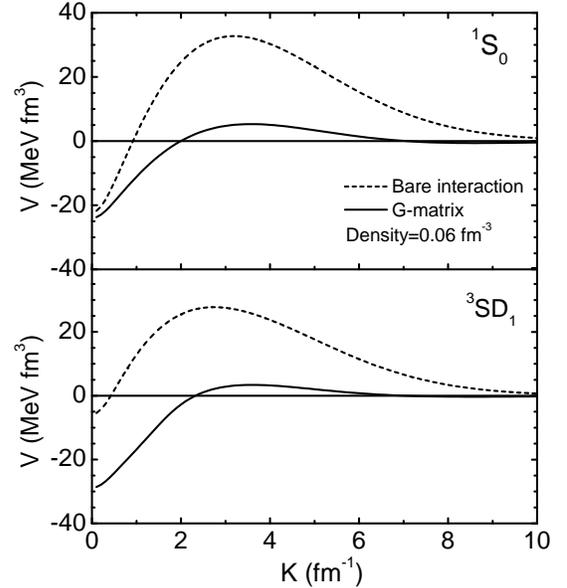}
 \vglue -4.0cm \caption{Shrinking of G-matrix in momentum space.}
\label{f:Gmat}
\end{figure}

 In a previous work the calculation of the screening
interaction was simplified by using the Gogny force, which in fact
reproduces most of the properties of a G-matrix. In the present
calculation we adopt the G-matrix itself and we try to reduce its
complexity with reasonable approximations. For the sake of
application to the pairing in the $^1S_0$ channel we select the
two particle state with total spin S=0 and isospin T=1. Then the
one-bubble interaction can be written as
\begin{widetext}
\begin{equation}
<1\bar 1|{\cal V}_1|1'\bar 1'> = \frac{1}{4}
\sum_{2,2'}\sum_{ST}(-)^S(2S+1)<12|G^{ph}_{ST}|1'2'>_A <2'\bar
1|G^{ph}_{ST}|2\bar1'>_A \Lambda^{0}(22'), \label{e:Vph}
\end{equation}
\end{widetext}
where $1\equiv(\vec k_1,\sigma_1,\tau_1)$ ($1'\equiv(\vec
k_{1'},\sigma_{1'},\tau_{1'}$)) and $\bar 1\equiv(-\vec
k_1,\sigma_1,\tau_1)$ ($\bar 1'\equiv(-\vec
k_{1'},\sigma_{1'},\tau_{1'})$) are the momenta of the pair in the
entrance (exit) channel. $\Lambda$ is the static polarization
part. The G-matrix is converted into the ph sector, as it is
required to solve the Bethe-Salpeter equation and to sum up the
bubble series $\tilde{V}_2$. The standard recoupling procedure
from pp sector to ph sector yields
\begin{widetext}
\begin{eqnarray}
G^{ph}_{ST}=\sum_{c}(2S_c+1)(2T_c+1)(-1)^{S_c+T_c}
\left\{%
\begin{array}{ccc}
  \frac{1}{2} & \frac{1}{2} & S_c \\
  \frac{1}{2} & \frac{1}{2} & S \\
\end{array}%
\right\}
\left\{%
\begin{array}{ccc}
  \frac{1}{2} & \frac{1}{2} & T_c \\
  \frac{1}{2} & \frac{1}{2} & T \\
\end{array}%
\right\} G_{S_cT_c}, \label{e: vpp}
\end{eqnarray}
\end{widetext}
where the brackets are the 6j symbols. The sum runs over the spin
$S_c$ and isospin $T_c$ of the pp channels included in the
calculation. Since the G-matrix incorporates short range pp
correlations, its momentum range is shrunk remarkably in
comparison with the bare interaction, as shown in
Fig.~\ref{f:Gmat}. At variance with the bare interaction, the the
G-matrix cannot sustain large momentum transfers $\vec q=\vec k -
\vec k'$, that justifies the approximation to average it around
the Fermi surface, in the limit q=0. As a consequence the q
dependence is only located in the integral of the polarization
part, giving the Lindhard function
\begin{eqnarray}
\sum_{\vec k} \Lambda^{0}_{\vec k,\vec k-\vec q}=
\frac{N(0)}{g}\frac{1}{2}\left[-1+\frac{1}{q}(1-\frac{q^2}{4})ln\left|
\frac{1-q/2}{1+q/2}\right|\right],   \label{lind}
\end{eqnarray}
where g is the degeneracy parameter.

The two external vertices mixing pp and ph lines shown in Fig. 1
(b) in principle induce unlimited excitations in momentum space.
But, since high momenta transitions are incorporated into G-matrix
used as vertex interaction, the main contribution of the
two-bubble diagram is concentrated in a domain as short as 2
 $fm^{-1}$. Elsewhere the remaining contribution can neglected with
respect to the bare interaction as shown in Fig.~2.

The problems with the calculation of the ph multi bubble
contribution (diagram (c) in Fig. 1) are the following. First, the
bubble series with G-matrix insertions have to be previously
summed up. But, since the interaction vertices in the ph channel
involve particle excitations around the Fermi surface, they can be
approximated by the Landau parameters. Second, even replacing the
bare interaction vertices by G-matrices , there appears the long
low density singularity in the RPA in nuclear matter ($F_0 = -1$).
This problem, discussed in Ref.~\cite{schul} (see also references
therein) is remedied by dressing the vertex insertions according
to the Babu-Brown induced interaction theory\cite{babu}.

\subsection{Landau parameters from the BHF approximation}

\begin{figure}[hbtp]
\includegraphics[scale=0.45]{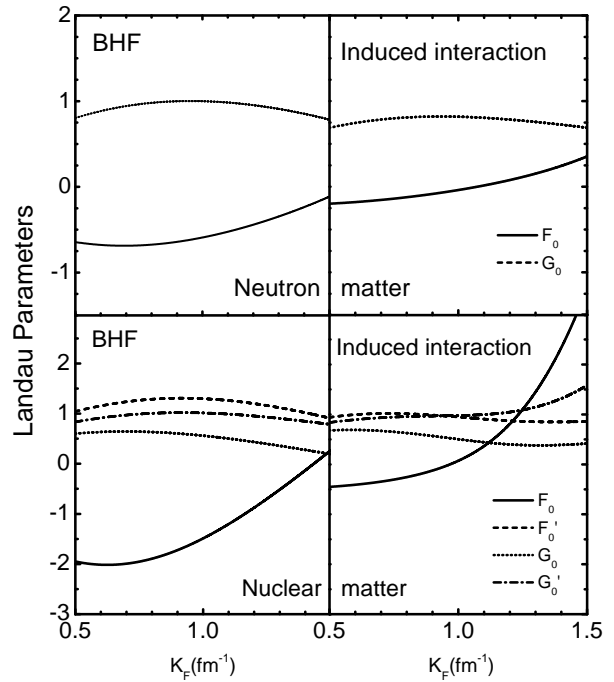}
\vglue -4.cm \caption{Landau parameters of pure neutron matter and
nuclear matter.} \label{f:landau}
\end{figure}

The microscopic basis of the ph effective interaction can be set in
terms of the energy functional of symmetric nuclear matter
%\begin{widetext}
\begin{eqnarray}
&N(0)f_{\sigma\tau,\sigma'\tau'}(\vec{k},\vec{k'})=\frac{\delta^2E}{\delta
n_{\sigma\tau}(\vec{k})\delta
n_{\sigma'\tau'}(\vec{k'})}& \\
  =&F + F' (\tau\cdot\tau') + G
(\sigma\cdot\sigma')+ G' (\sigma\cdot\sigma')(\tau\cdot\tau'),&
\nonumber \label{ph}
\end{eqnarray}
%\end{widetext}
where  the density of states $N(0)$ is introduced to make the
Landau parameters $F$, $F'$, $G$ and $G'$ dimensionless. In BHF
\cite{maha} approximation the energy functional is given by
\begin{eqnarray}
E=\sum_{k}\frac{\hbar^2k^2}{2m}+\frac{1}{2}\sum_{k_1,k_2}
\langle{k_1,k_2}|G(\omega)|k_1,k_2\rangle_A,
\end{eqnarray}
where the subscript $A$ means that the matrix element of the
G-matrix is antisymmetrized. The index $k$ stands for $\vec k$,
$\sigma$ and $\tau$, momentum, spin and isospin, respectively. The
G-matrix is understood to be calculated on the energy shell:
$\omega=\epsilon_{k_1}+\epsilon_{k_2}$. The single particle
energies are determined iteratively along with the G-matrix within
the Brueckner selfconsistent scheme. One can determine the Landau
parameters from the microscopic Brueckner theory in the BHF
approximation, performing the double variational derivative, Eq.
(8), of the energy per particle, Eq. (9). So doing, a number of
contributions are generated that can be calculated one by one
\cite{baldo} in some approximation due to the complex structure of
G-matrix. A simple and powerful way to calculate the Landau
parameters is to suitably  fit the BHF energy and the
corresponding sp spectrum with a functional of the occupation
numbers and then to perform the double derivative. A Skyrme-like
functional has proved to reproduce accurately the equation of
state (EoS) of symmetric as well as spin and isospin asymmetric
nuclear matter \cite{ligang}. Therefore we determine the Landau
parameters in that way. A limitation of this procedure is that
only a few partial wave components can be calculated, but for the
purpose of the present investigation we only need the zero order
Landau parameters. The latter are plotted in Fig.~\ref{f:landau}
as a function of the Fermi momentum. As expected $F_0$ exhibits
the well known instability below the saturation point, which makes
the RPA series difficult to handle. As in previous papers
\cite{hans,shen} this drawback can be overcome by the induced
interaction theory of Babu and Brown \cite{babu}. Leaving aside a
description of this theory (see Refs. \cite{back,sjob}), we
schematically write down the equation defining the ph induced
interaction as follows
\begin{eqnarray}
V_{ph} = V_d + {V}_{RPA}(V_{ph}).
\end{eqnarray}
The first term (direct term) is the BHF ph residual interaction,
which, in the first order, is represented by the G-matrix. The
second one (induced term) is the RPA bubble summation, in which
the vertex insertions are given by $V_{ph}$ itself instead of the
direct term.  The solution of the latter equation is quite simple
if we replace the true $V_d$ projection in the ph channel (ST)
with the corresponding Landau parameter, since the way we extract
the Landau parameters  the direct term contains not only the
effect of the G-matrix but also the rearrangement diagrams
\cite{back}. The numerical results are depicted in
Fig.~\ref{f:landau}. The salient feature of the induced
interaction is that the renormalization of $F_0$ prevents any
singular behavior to occur below the saturation density. Otherwise
the values do not differ from the Gogny interaction except for the
high density behavior where the effect of the three body force
makes $F_0$ much more repulsive than the Gogny force \cite{shen}.

Therefore we dressed the residual interaction first with the short
range correlations (G-matrix instead of bare interaction) and then
by the renormalized long range correlations $V_{ph}$ replacing the
G-matrix in the RPA series. Since the calculation of the induced
interaction with the G-matrix is a quite complex job, we have
simplified the problem replacing the G-matrix with the Landau
parameters. The way we determine the Landau parameters, the
approximation turns out to be better than starting from the
G-matrix itself.

\subsection{Bubble series}
The vertex insertions dressing the bubbles  must be treated on
different footing than the external ones (diagram  (c) of Fig. 1).
To sum up the RPA bubble series with the G-matrix requires to
solve the Bethe-Salpeter equation, what is a prohibitive task.
Using the Landau parameters corresponds to the Landau limit (zero
momentum-energy transfer around the Fermi surface), which is a
quite reasonable approximation. In this case the RPA summation of
the ph interaction turns out to be algebraic and, expressed in
term of the dressed bubble, it is written as
\begin{eqnarray}
\Lambda(q)_{ST} = \frac{\Lambda^0(q)}{1+\Lambda^0(q){\cal
L}_{ST}},
\end{eqnarray}
where ${\cal L}_{ST}$ are the Landau parameters, whose components
are commonly denoted by: ${\cal L}_{00}= F$, ${\cal L}_{01}=F'$,
${\cal L}_{10}=G$, ${\cal L}_{11}=G'$. In this expression we
clearly see how the induced interaction prevents any divergence to
occur since $|\Lambda{\cal L}|\leq|{\cal L}|\leq 1$. Replacing in
Eq.~(\ref{e:Vph}) the bare bubble $\Lambda^0$ with the dressed
bubble $\Lambda$ we get the full screening interaction used in the
calculation.

\section{Results}

The G-matrix is generated from a selfconsistent BHF calculation
with the continuous choice \cite{maha}. The Argonne V18 two body
force \cite{v18} is adopted as the input bare interaction plus a
microscopic model for the three body force based on meson exchange
with intermediate excitation of nucleon resonances (Delta, Roper,
and nucleon-antinucleon) \cite{tbf}. The calculation also provides
the self-energy from which we extract the sp spectrum and the
Z-factors. Based on the same framework is also the Skyrme-like fit
of the BHF energy functional used to calculate the Landau
parameters \cite{ligang}.

\subsection{Screening interaction}

In this paper we only focus on the $^1S_0$ pairing interaction in
the two extreme situations of pure neutron matter and symmetric
nuclear matter. We keep for a further investigation the
consideration of asymmetric nuclear matter with the purpose of
studying the transition from the screening regime in pure neutron
matter to the antiscreening regime in symmetric nuclear matter.

Let us start with neutron matter. In this case the screening
interaction can be decomposed in two terms: S=0 density fluctuation
and S=1 spin density fluctuation. The two modes have opposite
effect: the former one is attractive, the latter is repulsive. From
Eq.~(\ref{e:Vph}), we can write
\begin{eqnarray}
{\cal V}_1= \frac{1}{2}\Lambda(q)_{0}G_0^{ph} G_0^{ph} -
\frac{3}{2}\Lambda(q)_{1}G_1^{ph} G_1^{ph}.
\end{eqnarray}
The factor 3 is due to the multiplicity of the spin mode. For the
following discussion we should notice that $\Lambda$ is negative.
In turn, each G-matrix can be expressed as a superposition of
G-matrices, projected onto two particle states (see Eq.~(6)),
\begin{eqnarray}
G_0^{ph}& = & \frac{1}{2} (-G_{0} - 3G_{1}), \\
G_1^{ph}& = & \frac{1}{2} ( G_{0} -  G_{1}).
\end{eqnarray}
Since $G_0$ is attractive and $G_1$ is repulsive, assuming their
magnitude to be comparable, we get $G_0\approx G_1$
 \cite{heisel} and the multiplicity plays the main  role in establishing
the dominance of the spin density mode over the density mode. In
the latter calculation $G_0$ and  $G_1$ are only roughly
comparable, as it can bee seen in Fig.4, nevertheless the
conclusion is still valid. A quenching of $^1S_0$ pairing in
neutron matter is to be expected, a result established long ago
\cite{clark} and confirmed by many calculations in various
approximations (see \cite{schul} and references therein).

In nuclear matter the situation could be quite different since the
isospin fluctuations also come into play. The screening
interaction now is split according to Eq.~(\ref{e:Vph}) as follows
\begin{widetext}
\begin{eqnarray}
\nonumber
 {\cal V}_1= \frac{1}{4}(\Lambda(q)_{00}G_{00}^{ph}
G_{00}^{ph} + \Lambda(q)_{01}G_{01}^{ph} G_{01}^{ph})
-\frac{3}{4}(\Lambda(q)_{10}G_{10}^{ph} G_{10}^{ph}+
\Lambda(q)_{11}G_{11}^{ph} G_{11}^{ph}).\end{eqnarray}
\end{widetext}
The various contributions are plotted in Fig. 4 in terms of pp
states, the individual ph contributions are expressed as
\begin{eqnarray}
G_{00}^{ph} &=& \frac{1}{4} (G_{00} + 3G_{10}+ 3G_{01} + 9G_{11}), \\
G_{10}^{ph} &=& \frac{1}{4} (-G_{00}- 3G_{10}+  G_{01} + 3G_{11}), \\
G_{01}^{ph} &=& \frac{1}{4} (-G_{00}- 3G_{10}+  G_{01} + 3G_{11}), \\
G_{11}^{ph} &=& \frac{1}{4} ( G_{00}-  G_{10}-  G_{01} +  G_{11}).
\end{eqnarray}
In nuclear matter the pp G-matrix elements are dominated by the
deuteron channel ($^3SD_1$ coupled pp channel), which is very
attractive and therefore it reinforces the density mode and
weakens the spin mode. In other words, the main isospin effect is
to reverse the role of the medium, i.e. antiscreening instead of
screening. In previous papers this effect has been discussed in
terms of proton-proton ph screening against neutron-neutron ph
screening in the neutron-neutron $^1S_0$ channel \cite{sorb}. The
latter gives repulsion the former attraction. At variance with
Ref. \cite{heisel} the proton-proton ph screening is stronger than
neutron-neutron ph screening. This effect is to be traced back to
stronger in medium renormalization of the force in the $T=0$
channel than in the $T=0$ one. Antiscreening is the overall
effect.

In Fig. 5 we plot the full pairing interaction in the three
approximations used in the calculation of the energy gap. In
nuclear matter, as we discussed before, the screening effects in
fact reinforce the attractive strength of the bare interaction.
The main effect appears already at the one bubble level. The
deviation from the bare interaction increases at lower density. At
$k_F$ = 0.6 $fm^{-1}$ the enhancement is from -13 $MeV\cdot fm^3$
to -27 $MeV\cdot fm^3$. This is a huge variation which could
entail a large increase of the gap because it is exponentially
depending on the interaction. But at such a density the pair
correlations are rather weak  and thus we do not expect any large
increase of the gap. In the density domain of the maximum gap the
enhancement is much smaller and again we do not expect any
dramatic change in the gap magnitude as an effect of the
antiscreening. In Ref.~\cite{shen}, an improper coupling of the ph
states in the mixed representation prevented the cancelation among
different ph excitations to occur with the effect of producing a
more pronounced antiscreening.

 In neutron matter the
situation is the other way round. The screening is repulsive, and
small in the full RPA calculation, but still enough to produce a
sizeable quenching of the pairing gap. These predictions confirm
at least at qualitative level the corresponding results obtained
with the Gogny force \cite{shen}.

\begin{figure}[hbtp]
\includegraphics[scale=0.3]{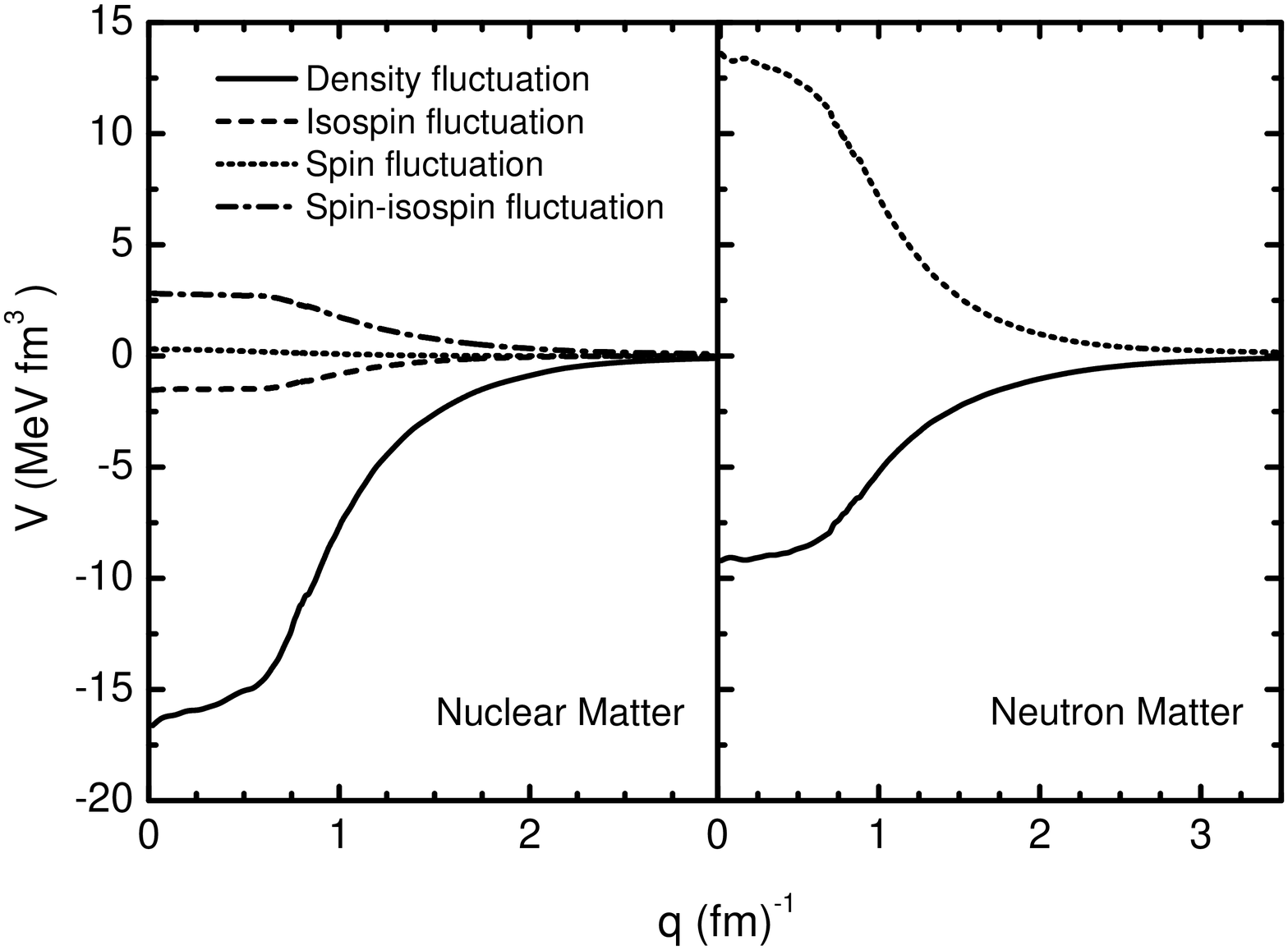}
\vglue -3.cm \caption{Individual components of the ph residual
interaction.} \label{f:Vph}
\end{figure}

\begin{figure}[hbtp]

\includegraphics[scale=0.3]{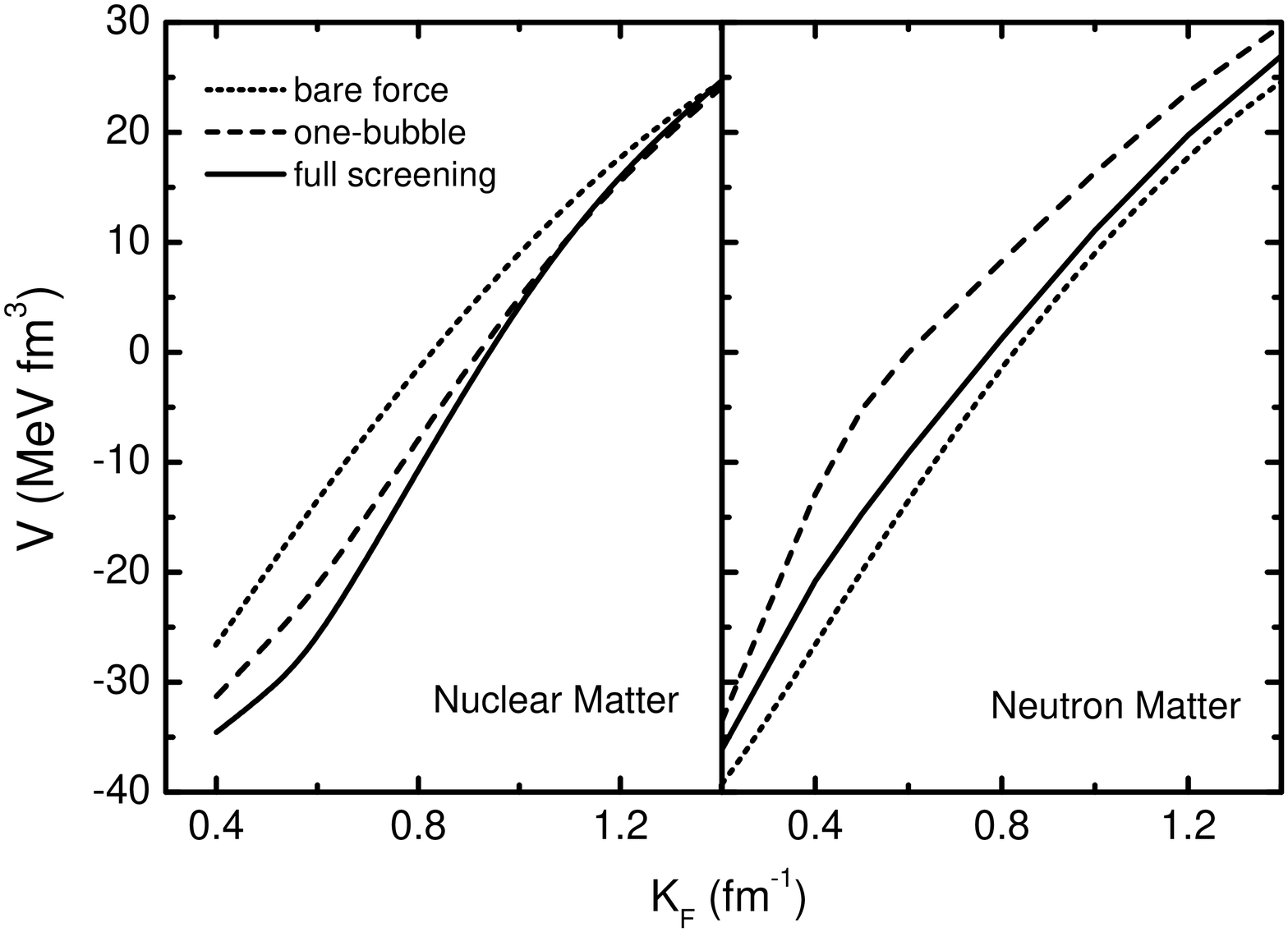}
\vglue -3.cm \caption{Pairing interaction.} \label{f:full}
\end{figure}

\subsection{Pairing gap}

The present calculation is  focussed on the $^1S_0$
neutron-neutron (or proton-proton) pairing. One can distinguish
the bare interaction which is responsible for the pairing between
the two particles in the $^1S_0$ state, from the screening
interaction induced by the surrounding particles. Therefore the
interaction, projected onto the $^1S_0$ channel, can be cast as
follows
\begin{equation}
<k|{\cal V}|k'> = \int \frac{d\Omega}{4\pi} [\mathcal{V}_0(\vec
k,\vec k') + \mathcal{V}_1(|\vec k - \vec k'|)].
\end{equation}

As bare force we use Argonne $V18$, the same as for calculating the
G-matrix and the selfenergy. The BCS energy gap in the $^1S_0$
channel is practically independent of the adopted bare force, since
in fact all realistic interactions reproduce the phase shifts of
free NN scattering.

\begin{figure}[hbtp]
\includegraphics[scale=0.4]{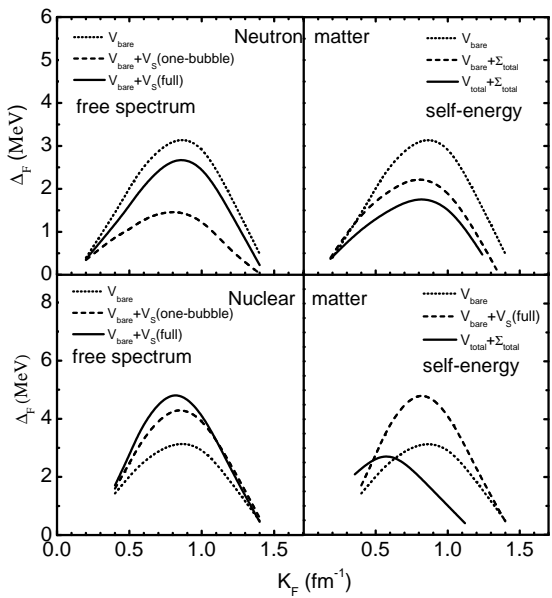}
\vglue -4.cm \caption{Pairing gap in the $^1S_0$ channel for pure
neutron matter (upper figure) and nuclear matter (lower figure).}
\label{e:gapv}
\end{figure}

We solved the gap equation in the form of Eq.~(\ref{e:zzgap}). In
order to disentangle the screening effects from the selfenergy
corrections, we first assume Z=1 and free sp spectrum. The results
are plotted in Fig. \ref{e:gapv} (upper left panel). In neutron
matter the screening effect is small and just reduces the gap by
$10\%$ in the peak region. At variance with previous calculations
existing in the literature \cite{schul} the full RPA screening is
much less effective than the one bubble approximation because of
the stronger renormalization of the spin fluctuations vs the
density fluctuations in the induced interaction. However this
finding confirms the preceding predictions with Gogny force (see
Fig.8 of Ref.~\cite{shen}).

In nuclear matter, due to the antiscreening effect we discussed
earlier, the magnitude of the gap variation is the other way
around and much more sizeable: the gap rises up from 3 $MeV$ to 5
$MeV$ for Fermi momentum $k_F$ = 0.8 $fm^{-1}$. This is displayed
in Fig. 6 (lower left panel).

There are two kinds of selfenergy effects: dispersive effect and
Fermi surface depletion. Both are calculated taking into account
the selfenergy corrections at the second order of G-matrix
(rearrangement terms). The first one is a correction to the sp
spectrum in the energy denominator. Usually it entails a reduction
of the pairing gap since the effective mass, beyond BHF
approximation, is less than the unity (the effective mass is the
combination of the e-mass and the k-mass \cite{maha}). But at very
low density the effective mass is larger than unity \cite{zuo} and
it reduces the quenching rate of the gap due to the interaction.
This effect can be seen in the low density side of the neutron gap
with $\Sigma_{total}$ (upper right panel). Additional strong
reduction is due to the depletion of the Fermi surface which
hinders transitions around the Fermi surface. The maximum gap in a
complete calculation is $1.5 - 2$ $MeV$ at $k_F\approx$ 0.8
$fm^{-1}$.

In nuclear matter the selfenergy effects are much stronger already
at moderately low density, as it has to be expected, and the peak
value shifts down to very low density $k_F\approx  0.5 - 0.6$
$fm^{-1}$. The Z-factor plays the major role: it quenches from
0.84 in neutron matter to 0.68 in nuclear matter at $k_F$ = 0.8
$fm^{-1}$. But the magnitude is about 0.5 $MeV$ less than the
value with only bare interaction. Therefore we can conclude that a
strong cancelation occurs as soon as  vertex corrections and self
energy effects are simultaneously included in the gap equation.
But this happens only in nuclear matter as an effect of
antiscreening. We will come back to this point below.

\section{Discussion and conclusions}

In this paper an exhaustive treatment of the $^1S_0$ pairing in
nuclear and neutron matter has been reported. The medium
polarization effects on the interaction and the selfenergy
corrections to the mean field, both developed in the framework of
the Brueckner theory, have been included in the solution of the
gap equation.

Within the pure mean field approximation \cite{liege} the $^1S_0$
gap is not affected by the medium, either nuclear or neutron
matter. So far the medium effects have not been considered in the
case of nuclear matter except in Ref.~\cite{shen}. The vertex
corrections due to neutron matter all give a reduction of the
pairing, the magnitude depending on the adopted approximation
\cite{schul}. The explanation relies on the competition of the
attractive density excitations against the repulsive spin density
excitations. The present calculation, based on G-matrix, also
predicts a large quenching in agreement with almost all previous
predictions, but only at the one bubble level. In the most
complete calculation (full RPA) the quenching is largely reduced
in apparent agreement with a recent Monte Carlo calculation
\cite{fantoni}. But the inclusion of selfenergy effects definitely
results in a large suppression as expected from basic properties
of a strongly correlated many body system (see Introduction).

In the case of nuclear matter the most remarkable result is the
antiscreening effect of the medium polarization. In fact in
nuclear matter isospin modes arise that reverts the competition
between the attractive density modes and the  repulsive
spin-density modes due to the presence of isospin modes. The
argument addressed in Ref.~\cite{heisel} that the p-n (T=0)
interaction is small compared to the n-n (T=1) is based on the
vacuum scattering T-matrix and does not consider the strong medium
renormalization of G-matrix, which inverts the strength of the two
channels. However the enhancement of the gap to almost 5 MeV is
almost completely suppressed by the strong correlation effects on
the selfenergy. But, even a small variation of the force strength
implies a large variation of the gap. These effects also push to
lower density the peak value of the gap.

Calculations of the pairing gap in a nuclear environment have been
reported in a series of papers for the case of nuclei
\cite{milan}. Their main finding is that the induced interaction
arising from the surface vibrations is responsible for large part
of the experimental gap. This result can be considered as the
counterpart for finite nuclei of the antiscreening effect due to
the medium polarization in nuclear matter. But the selfenergy
effects completely compensate the gap enhancement and, in the end,
the full medium effect do not change significantly the gap with
bare interaction. This result turns out to be not a big surprise,
since some calculations show that the gap with Gogny interaction
is consistent with the observed gaps in nuclei \cite{schuck}.

At this point two aspect are worth to be developed further. The
first one is the study of pairing in the transition from symmetric
nuclear matter to neutron matter; the second one is the
investigation of pp channels so far neglected, since there the
bare interaction is repulsive. The pairing could exist as an
induced effect of the environment.

\subsection*{Acknowledgments}

This work was partially supported by the grant appointed to the
European Community project {\it Asia-Europe Link in Nuclear Physics
and Astrophysics}, CN/ASIA-LINK/008(94791).

\end{document}